\documentclass[aps,twocolumn,showpacs]{revtex4}
\usepackage{amsmath}
\usepackage{graphicx}

\begin{document}

\title{A Model for Conductive Percolation in Ordered Nanowire Arrays}
\author{J. L. Silverberg}
\email{silverberg.je@neu.edu}
\affiliation{Physics Department, Northeastern University,}
\affiliation{Department of Mathematics, Northeastern University, Boston MA 02115}

\begin{abstract}
The combined processes of anodization and electrodeposition lead to highly ordered arrays of cylindrical nanowires.  This template-based self-assembly fabrication method yields nanowires embedded in alumina.  Commonly, chemical etching is used to remove the alumina and free the nanowires.  However, it has been experimentally observed during the etching process that the nanowires tend to form clumps.  In this work, the nanowires are modeled as elastic rods subject to surface interaction forces.  The dynamics of the model give rise to the aforementioned clumping behavior which is studied via percolation theory.  This work finds that percolation takes place with probability $P \sim (t-t_c)^x$, where the exponent $x = 2.8$ and $t_c$ is the time at which percolation takes place.  The critical exponents which entirely determine the system are found to be for (dimension) $d = 2$, $\beta = 2.1, \ \gamma = 0.57, \ \Delta = 2.7, \ \alpha = -2.8, \ \nu = 2.4,$ and $\delta = 1.3$.
\end{abstract}
\maketitle

\section{\label{intro} Introduction}
The fabrication of nanowires through anodization and electrodeposition provides an interesting opportunity for the study of percolation.  Before presenting the details of how this comes about, we first introduce the related key topics.

\subsection{Template Based Fabrication}
Under proper conditions, the anodization of aluminum creates a thin layer of alumina containing a highly ordered hcp array of pores \cite{Evans:2006}.  These pores, though typically of a uniform size in a well made sample, can have diameters ranging from 10 nm to 100 nm.  Their length, being a function of time, can easily extend to the micron scale.  Electrodeposition of metals such as Au, Ag, Co, Ni, or Fe results in a self-assembled and highly ordered array of cylindrical nanowires.

\begin{figure}
\scalebox {0.45}{\includegraphics{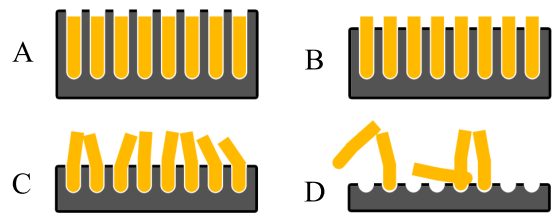}}
\caption{Seen from a cross sectional slice, the anodized alumina acts as a template for nanowire growth.  The sequence A-D illustrates various etch times where A is no etching; B is moderate etching which exposes just the tips; C is extended etching which results in clumps of nanowires; and D is total etching which releases the nanowires.  Region C is of particular interest because this is when the percolation phenomena begins to takes place.}
\label{Fig1}
\end{figure}

Post-fabrication, the nanowires exist as a bulk sample; they remain embedded within the alumina.  If it is so desired, a chemical etch can remove the alumina and release the nanowires into solution.  Figure \ref{Fig1} illustrates the etching of a sample at different times.  

\subsection{Conductive Percolation}
Percolation theory has been used to study conductivity in a variety of electrical systems \cite{Stauffer:1987, Essam:1980}, but more recently has seen applications to carbon nanotubes \cite{Snow:2003, Kumar:2005} and the growth of V$_2$O$_5$ nanowires \cite{Chang:2004}.  In these examples, the nanostructures are initially suspended in solution and allowed to settle on the surface of a substrate.  The resulting random network is probed for a conductive path as a function of surface coverage.  Generally, it is found that at some critical density, the conductivity rapidly increases as more electrical paths become available for the current.  

\subsection{Clumping of Nanowires}
\begin{figure}
\scalebox {.50}{\includegraphics{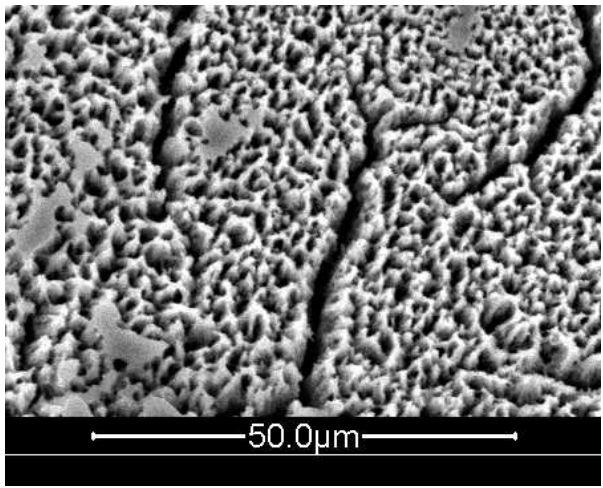}}
\caption{During extended etch times, the exposed portions of the nanowires bend from their initial position and clump into groups.  This results in large clusters of nanowires making physical contact and leads to the possibility of conductive percolation.  In this SEM image, the etching has taken place for sufficiently long time.  Clumps have begun to take shape as the nanowires bend due to surface forces.}
\label{Fig2}
\end{figure}

The key observation which relates percolation and  nanowire fabrication in alumina templates, is the behavior of clumping.  As has been reported \cite{Silverberg:2007, Evans:2006, Rabin:2003}, during the etching process, nanowires tend to clump into bundles.  If the process is interrupted at various times, the average clump size (i.e. number of nanowires in direct physical contact) is seen to grow.  Figure \ref{Fig2} shows a typical example of this behavior.  

In an experimentally realizable scenario, the clumped nanowires can be tested for conductive percolation by placing electrical contacts along two opposing edges of the sample.  When the template has been sufficiently etched (as in figure \ref{Fig1}C), an electrical path should be observed in the form of a percolating cluster.    

Inspired by these results, we consider a two dimensional array of nanowires, modeled as elastic rods, and subject to van der Waals interaction forces.  In the model, we consider the nanowire length a linear function of time to simulate the effect of etching.  The dynamics of this system are simulated in MATLAB and it is observed that given sufficient etch time, the nanowires bend in the direction of their neighbors.  This gives rise to clusters of variable size which we interpret from the view of percolation theory.  In section \ref{surface} the van der Waals interaction force is derived for the cylindrical geometry of two nanowires.  In section \ref{rod} we review the equations of equilibrium which describe the bending of elastic rods.  In section \ref{num} the numerical results are presented and discussed in terms of percolation.  In section \ref{conc} two experiments are proposed which would quantitatively verify the model described in this work.

\section{\label{surface}Surface Interactions}
\begin{figure}
\scalebox {1.25}{\includegraphics{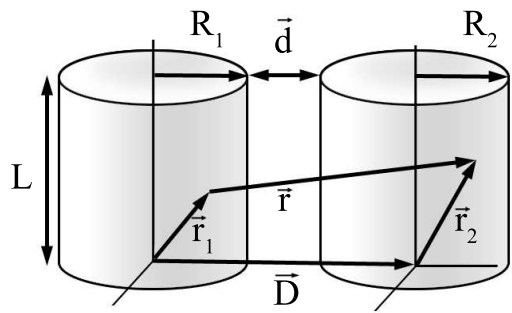}}
\caption{In calculating the van der Waals interaction energy, we use the geometry of two cylinders of equal length $L$, separation $d$, radius $R_1$ and $R_2$.  The vectors in Eqs.(\ref{surface3}) are defined as shown.}
\label{Fig3}
\end{figure}

To calculate the interaction forces between two macroscopic bodies, we use the non-retarded van der Waals potential.  If we consider the bodies to be composed of discrete point-like atoms, the total energy of interaction can be written as the sum 
\begin{equation}
\label{surface1}
U = \sum_{i,j} U_{ij} = - C \sum_{i,j} r_{ij}^{-6}.
\end{equation}
We denote by $r_{ij}$ the separation between two given particles, the $i^{\rm th}$ in the first body and the $j^{\rm th}$ in the second.  The constant $C$ is the coefficient of the atom-atom pair potential.  Passing to the continuum limit,
\begin{equation}
\label{surface2}
U = -C \varrho_1 \varrho_2 \int_{V_1}\int_{V_2}\frac{dV_1 dV_2}{|{\bf r}|^6},
\end{equation}
where $\varrho$ is the number of atoms per unit volume.  Because we are interested in the interaction of two nanowires, the natural coordinates are cylindrical (see figure \ref{Fig3}).  It can be seen from simple vector algebra that
\begin{eqnarray}
\label{surface3}
{\bf D} & = & \left\langle R_1 + R_2 + d , 0 , 0 \right\rangle, \nonumber \\
{\bf r}_1 & = & \left\langle \rho_1 \cos \theta_1 , \rho_1 \sin \theta_1 , z_1 \right\rangle, \nonumber \\
{\bf r}_2 & = & \left\langle \rho_2 \cos \theta_2 , \rho_2 \sin \theta_2 , z_2 \right\rangle, \ \ \ {\rm and} \nonumber \\
{\bf r} & = & {\bf D} + {\bf r}_2 - {\bf r}_1. 
\end{eqnarray}
Expanding the denominator of Eq.(\ref{surface2}) in terms of components, the integral can be calculated.  The final result is \cite{Israel:1991}
\begin{equation}
\label{surface4}
U({\bf d}) = -\frac{AL}{12\sqrt{2} \ |{\bf d}|^{3/2}}\left( \frac{R_1 R_2}{R_1 + R_2} \right)^{1/2},
\end{equation}
where A is the Hamaker constant defined by 
\begin{equation}
\label{surface4a}
A = \pi^2 C \varrho_1 \varrho_2.
\end{equation}

The force arising from this potential is,
\begin{equation}
\label{surface5}
{\bf F}({\bf d}) = - {\bf \hat{n}} \frac{AL}{8\sqrt{2} \ |{\bf d}|^{5/2}} \left( \frac{R_1 R_2}{R_1 + R_2} \right)^{1/2},
\end{equation}
where ${\bf \hat{n}} = {\bf d} / d$.  Letting $R_1 = R_2 \equiv R$, 
\begin{equation}
\label{surface6}
{\bf F}({\bf d}) = - {\bf \hat{n}} \frac{AL}{16} \left(\frac{R}{|{\bf d}|^5}\right)^{1/2}.
\end{equation}
When immersed in aqueous solution, the Hamaker constant is not accurately given by Eq.(\ref{surface4a}).  Because interactions in the medium effect the pair-wise potential, experimentally determined results for metals in water must be used \cite{Derjaguin:1978}.

\section{\label{rod}Equilibrium of a Bending Rod}
We now review the equations of equilibrium for a rod subject to external forces.  The discussion closely follows \cite{LandauTE:1998} and only the main results will be presented.

Consider a thin rod of circular cross-section.  To an infinitesimal length, assign local coordinates $\xi$, $\eta$, $\zeta$ so the $\zeta$-axis is aligned parallel to the rod.  Furthermore, let ${\bf r} = {\bf r}(l)$ be a curve parameterized by arc length describing the local coordinates, and hence, the unit tangent vector is given by ${\bf t} = d{\bf r}/dl$.  We define
\begin{equation}
\label{rod1}
{\bf \Omega} = d{\bf \Phi}/ dl,
\end{equation}
to be the rotation of the local coordinates along the rod and notice that in the absence of torsion, ${\bf t} \cdot {\bf \Omega} = \Omega_{\zeta} = 0$.  Moving along an infinitesimal length, the change in the tangent vector is given by
\begin{equation}
\label{rod2}
\frac{d {\bf t}}{dl} = \frac{d {\bf \Phi}}{d l} \times {\bf t} = {\bf \Omega} \times {\bf t}.
\end{equation}
Taking the cross product of both sides with ${\bf t}$ and rearranging, we find the expression
\begin{equation}
\label{rod3}
{\bf \Omega} = {\bf t} \times \frac{d {\bf t}}{dl} + {\bf t}({\bf t} \cdot {\bf \Omega}).
\end{equation}
In the case of pure bending, the last term is zero due to the absence of torsion.  Let the moment vector $\bf M$ be defined by
\begin{equation}
\label{rod4}
{\bf M} = E I {\bf \Omega},
\end{equation}
where $E$ is Young's modulus and $I$ is the moment of inertia of the cross section for a cylinder.  As a condition for equilibrium, the total moment of the forces is zero, i.e., $d {\bf M} + d{\bf r}\times {\bf F} = 0$.  Dividing by $dl$,
\begin{equation}
\label{rod5}
d{\bf M} / d{\bf l} = {\bf F} \times {\bf t}.
\end{equation}
Rearranging Eqs.(\ref{rod3} - \ref{rod5}), we find the equation of equilibrium for pure bending in a circular rod:
\begin{equation}
\label{rod6}
{\bf F} \times \frac{d{\bf r}}{d l} = E I \frac{d {\bf r}}{d l} \times \frac{d^3 {\bf r}}{d l^3}.
\end{equation}

\section{\label{num}Numerical Results}
To estimate the magnitude of the force in Eq.(\ref{surface6}) and the amount of displacement the tip of a nanowire will undergo, characteristic values are used \cite{Evans:2006, Silverberg:2007, Derjaguin:1978, Huang:2002, Yin:2001}.  For Au nanowires with radius $R = 20$ nm, interpore separation $d = 10$ nm, length $L = 500$ nm, and Hamaker constant $A = 10^{-20}$ J, the force $F = 44$ pN.  The amount of deflection can be found by treating the nanowire as a cantilever with deflection $\Delta d = F L^3 / 3 E I$.  For Au, Young's modulus $E = 7.8 \times 10^7$ Pa, the moment of inertia of the cross-section is $I = (\pi/4)R^4 = 1.3 \times 10^{-31}$ m$^4$, and consequently, the deflection is $\Delta d = 19$ nm.  This estimate indicates that van der Waals forces are sufficiently capable of bending the nanowires.  Consequently, we proceed to a more sophisticated numerical solution of the equations involving an array of nanowires.

\begin{figure}
\scalebox {.45}{\includegraphics{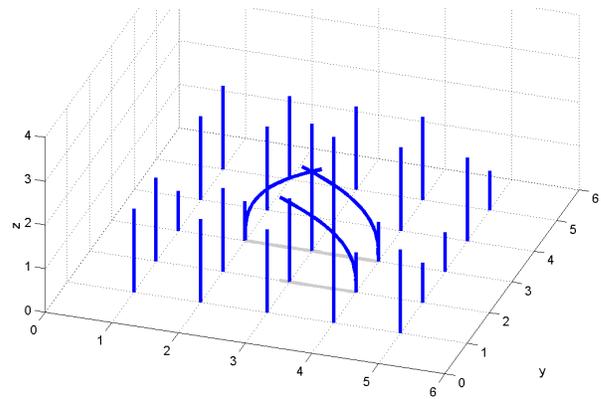}}
\caption{A $5 \times 5$ grid is simulated for $t_{max} = 3$ to illustrate the results of the model.  As shown, some nanowires are already bending to make contact with their neighbors while others remain unaffected.  For computational reasons, the perimeter is considered infinitely rigid and those wires do not bend.}
\label{Fig4}
\end{figure}

The array of nanowires is simulated using a $4^{\rm th}$ order Runge-Kutta method to solve Eq.(\ref{rod6}) subject to the forces given by Eq.(\ref{surface6}) on a square grid.  Each grid site $\{ij\}$ corresponds to an individual nanowire and the forces acting on it are determined by nearest neighbor interactions solely concentrated at the tip.  Because each nanowire is treated as an elastic rod, the net forces cause bending accordingly.  The continuous etching of alumina shown in figure \ref{Fig1} is simulated by making the length of each nanowire a linear function of time.  Thus, taking $z_{ij} = v_0 t$ where $t \in (0,t_{max}]$ and $v_0$ is constant, the process can be stopped at each unit of time and checked for percolation.  Furthermore, the length of each nanowire is randomized to within $15\%$ of the maximum value to reflect experimental observations.  The boundries were treated as infinitely stiff rods and hence, not subject to interaction forces.  Figure \ref{Fig4} shows a $5 \times 5$ sample simulation where some nanowires have already made physical contact with their neighbors.

\begin{figure}
\scalebox {.85}{\includegraphics{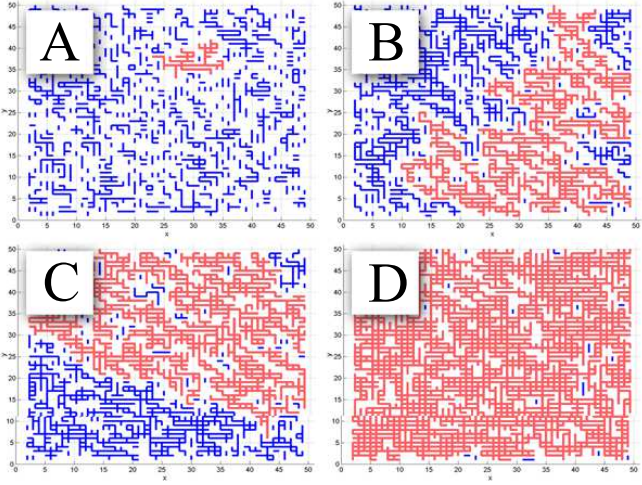}}
\caption{Typical results for simulations of various times.  (A) $t = 3$; (B) $t = 6$; (C) $t = 8$; and (D) $t = 13$.  Blue line segments indicate two nanowires that are touching.  Blank sites indicate free-standing nanowires, and the largest cluster in each figure was colored red to illustrate the growth in cluster size as a function of time.  Both (C) and (D) contain percolating clusters.}
\label{Fig5}
\end{figure}

\begin{figure}
\scalebox {.55}{\includegraphics{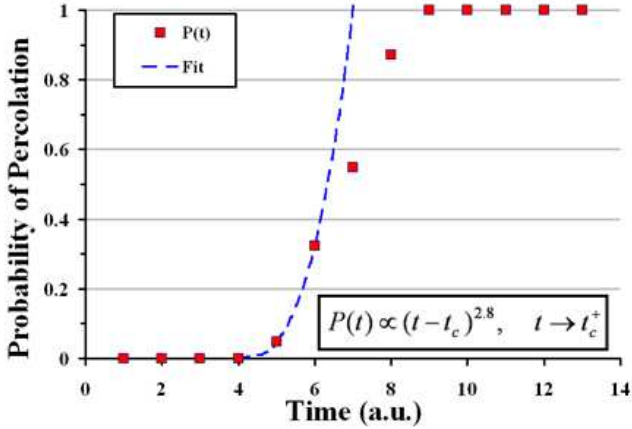}}
\caption{The probability of percolation is plotted against the simulation etch time.  It is seen that at $t \approx 4$ the probability grows like $P \propto (t-t_{c})^{x},$ where $t_c = 4$ and the critical exponent $x = 2.8$.}
\label{Fig6}
\end{figure}

In a $2D$ representation, figure \ref{Fig5} shows typical results for repeated simulations of various etch times.  The blue line segments represent physical contact between two grid points and the largest clusters have been colored red.  Each simulation was deemed percolating when a cluster spanned the full horizontal distance.  Running 500 simulations for each $t_{max} = 1,2,...,13$, the probability of percolation was determined.  Figure \ref{Fig6} displays these results along with the fitted curve $P \propto (t-t_{c})^{x},$ where $t_c = 4$ and the critical exponent $x = 2.8$.  This nonlinear scaling is indicative of the percolation threshold.  

\begin{figure}
\scalebox {.55}{\includegraphics{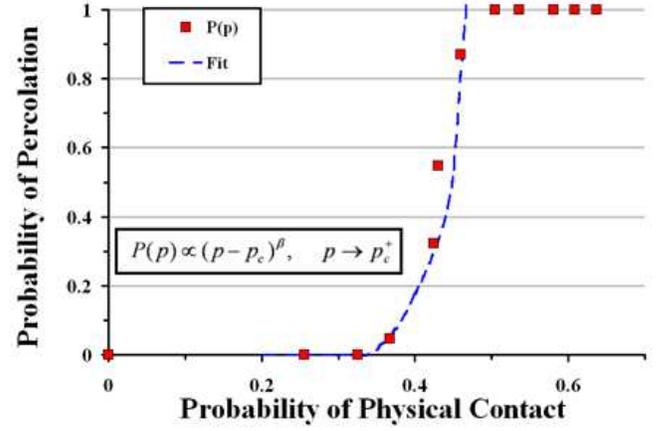}}
\caption{The probability of percolation is plotted against the probability of physical contact between two wires.  This scales like $(p-p_c)^\beta, p \rightarrow p_c^+$, which aids in determining the full set of critical exponents.  From the fit, it is observed that $\beta = 2.1$.}
\label{Fig7}
\end{figure}

\begin{figure}
\scalebox {.55}{\includegraphics{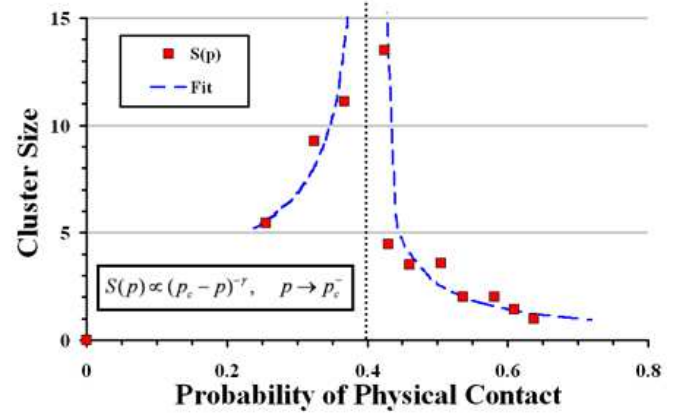}}
\caption{The cluster size is plotted against the probability of physical contact between two wires.  The singularity in the neighborhood of $p \approx 0.4$, is due to the presence of large clusters.  As the threshold is crossed and the clusters become infinite in size, they make no contributions to the average cluster size and consequently, the curve decays as $(p_c - p)^{-\gamma}, p \rightarrow p_c^-$.  From the fit, it is observed that $\gamma = 0.57$. }
\label{Fig8}
\end{figure}

Near the threshold, percolation theory predicts that the probability of percolation as a function of the probability of physical contact, $P(p)$, will obey \cite{Stauffer:1987}
\begin{equation}
\label{num1}
P \propto (p-p_c)^\beta, \ \ \ p \rightarrow p_c^+,
\end{equation}
where $\beta$ can be related to other critical exponents of the system.  Similarly, the mean cluster size as a function of the probability of physical contact, $S(p)$, goes like 
\begin{equation}
\label{num2}
S \propto (p_c-p)^{-\gamma}, \ \ \ p \rightarrow p_c^-.
\end{equation}
From figures \ref{Fig7} and \ref{Fig8}, $\beta = 2.1$ and $\gamma = 0.57$, can be read from the curve fits. The full set of critical exponents is determined by the equations
\begin{eqnarray}
\Delta & = & \gamma + \beta, \nonumber \\
\alpha & = & 2 - \Delta - \beta, \nonumber \\
\nu & = & \frac{2 - \alpha}{2}, \ \ \ {\rm and} \nonumber \\
\frac{1}{\delta} & = & \frac{2 - \alpha}{\Delta} - 1.
\label{num3}
\end{eqnarray}
Therefore, in (dimension) $d = 2$ for the percolating nanowire network subject to van der Waals interaction, the complete set of critical exponents is given by $\beta = 2.1, \ \gamma = 0.57,$ and with the aid of Eqs.(\ref{num3}), $\Delta = 2.7, \ \alpha = -2.8, \ \nu = 2.4,$ and $\delta = 1.3$.

\section{\label{conc}Conclusion}
The goal of this work was to quantitatively explain and describe the phenomenon of nanowire clumping during the etching of ordered nanowire arrays.  Having accomplished this task, we now outline two potential experiments which could be used to verify the predicted critical exponents.  In the first case, a number of samples could be fabricated in the laboratory and by means of \textit{counting small subregions}, all the numerically simulated results could be readily verified.  

The alternative experiment involves measuring the conductivity across an array of nanowires while the alumina is etched.  At time $t = 0$, all the nanowires would be embedded in the alumina (an insulator) and no current would flow.  At some time $t = t_c$, the first percolating cluster would form and the current density would increase from $0$.  Continuing the measurements for longer time, the conductivity would increase further as more pathways through the nanowire array opened up allowing a greater current to flow.  As $t \rightarrow \infty$, the conductivity would approach a constant value when the clusters have fully formed and no more potential pathways exist.  By means of the usual analysis in conductive percolation \cite{Kirkpatrick:1973}, the relevant statistical distributions can be determined from this data.  

In conclusion, the qualitative observation of clumping was modeled by elastic rods subject to van der Waals interaction forces.  Simulations in MATLAB revealed a percolative phase transition, from which, the critical exponents were extracted.  They are $\beta = 2.1, \ \gamma = 0.57, \ \Delta = 2.7, \ \alpha = -2.8, \ \nu = 2.4,$ and $\delta = 1.3$.

\begin{acknowledgments}
J. L. S. would like to thank M. Schwarz and L. Menon for their support, as well as A. Ghosh and A. Singh for their helpful discussions in this work.  
\end{acknowledgments}

\end{document}